\documentstyle[psfig]{l-aa-lett} 
\def\pBoxtimes{{\ooalign{\hfil\raise.2ex\relax
\hbox{$\times$}\hfil\crcr\hbox{$\Box$}}}}

\def\pBoxcirc{{\ooalign{\hfil\raise.2ex\relax
\hbox{$\circ$}\hfil\crcr\hbox{$\Box$}}}}

\def\pDiatimes{{\ooalign{\hfil\raise.2ex\relax
\hbox{$\times$}\hfil\crcr\hbox{$\Diamond$}}}}

\def\pBoxplus{{\ooalign{\hfil\raise.2ex\relax
\hbox{$+$}\hfil\crcr\hbox{$\Box$}}}}
\newcommand{\aj}{AJ}         
\newcommand{\aaa}{A\&A}      
\newcommand{\aas}{A\&AS}     
\newcommand{\apj}{ApJ}       
\newcommand{\apjs}{ApJS}     
\newcommand{\mnras}{MNRAS}   
\newcommand{\pasp}{PASP}     
 
\newcommand{\etal}{et al.\ }

\newcommand{\eg}{e.g.,\ }

\newcommand{\kms}{\mbox{km s$^{-1}$}}

\newcommand{\Ha}{H$\alpha$}
\newcommand{\Hb}{H$\beta$}

\newcommand{\OI}{[{\sc O$\,$i}]}
\newcommand{\OII}{[{\sc O$\,$ii}]}
\newcommand{\OIII}{[{\sc O$\,$iii}]}
\newcommand{\NI}{[{\sc N$\,$i}]}
\newcommand{\NII}{[{\sc N$\,$ii}]}
\newcommand{\NeIII}{[{\sc Ne$\,$iii}]}
\newcommand{\SII}{[{\sc S$\,$ii}]}

\newcommand{\multi}{\multicolumn}

\def\mgb{Mg\,$b$} 

\begin{document}

\title{Ionized gas in early-type galaxies:\ Its effect on Mg\,$b$ and other
stellar line-strength indices\thanks{Partly based on observations obtained at
the European Southern Observatory, La Silla, Chile, and at the
Canada-France-Hawaii Telescope, operated by the National Research Council of
Canada, the Centre National de la Recherche Scientifique of France, and the
University of Hawaii}}

\author{Paul Goudfrooij\inst{1,}\thanks{E-mail:\ pgoudfro@eso.org}
\and Eric Emsellem\inst{2,}\thanks{E-mail:\ emsellem@strw.LeidenUniv.nl}} 

\offprints{Paul Goudfrooij}

\institute{
 European Southern Observatory, Karl\--Schwarz\-schild\--Strasse 2,
 D-85748 Gar\-ching bei M\"un\-chen, Germany 
\and 
 Sterrewacht Leiden, Postbus 9513, NL-2300 RA Leiden, The Netherlands 
}

\thesaurus{03(03.20.8; 11.05.1; 11.09.1 NGC 2974; 11.09.4; 11.19.6)}

\date{Received December 7, 1995 / Accepted January 11, 1996}

\maketitle

\markboth{P.\ Goudfrooij \& E.\ Emsellem:\ Ionized gas and stellar
line-strength indices in galaxies}
{P.\ Goudfrooij \& E.\ Emsellem:\ Ionized gas and stellar
line-strength indices in galaxies} 

\begin{abstract}
In the light of the recent finding that nebular emission lines are 
commonly found in the inner regions of early-type galaxies, we evaluate their
effect to stellar absorption-line indices. We derive analytical expressions 
for changes induced by the presence of nebular emission lines, both
for atomic and molecular absorption-line indices. 
We find that the \NI\ emission-line doublet at 5199 \AA\ can
significantly affect \mgb\ line-strength measurements. For typical
equivalent widths of the \NI\ doublet in nuclei of early-type
galaxies featuring nebular emission, the \mgb\ index is artificially
enhanced by 0.4\,--\,2 \AA, which represents a significant fraction of
the  typical equivalent width of the \mgb\ line in early-type galaxies.  

We illustrate this effect in the case of NGC~2974.


\end{abstract}

\section{Introduction}
\label{intro} 

The \mgb\ index at 5177\,\AA\ is one of the most prominent features in the
optical spectra of old stars. 
It is one of the 11 well-known strong spectral line-strength indices that are
widely used to study stellar populations in old stellar systems such as
globular clusters and early-type galaxies (e.g., Burstein \etal \cite{bfgk84};
Faber \etal \cite{fab+85}; Gorgas, Efstathiou \& Arag\'on-Salamanca
\cite{gea90}; Davies, Sadler \& Peletier \cite{dsp93};
Gonz\'alez \cite{gonz93}; Carollo \& Danziger
\cite{cardan94};  Sansom, Peace \& Dodd \cite{sams+94}; Fisher, Franx \&
Illingworth \cite{fish+95}; Surma \& Bender \cite{surben95}).  
Using a library of standard stars, these indices have been empirically
calibrated as a function of stellar colour, surface gravity and metallicity
(Gorgas \etal \cite{gorg+93}), allowing the construction of semi-empirical
population synthesis models (e.g., Worthey \cite{guy94}). The
behaviour of these indices in old stellar systems has provided useful
insights on the star formation history of elliptical galaxies (e.g.,
Davies \cite{dav95} and references therein). 

These line strengths are a measure of the depth of a given
absorption feature, based on the flux in a central bandpass covering the
feature of interest versus a local continuum level, which is linearly
interpolated from a pair of bracketing bandpasses at either side of the
feature. The continuum bandpasses have been defined in regions close to the
feature, having no strong absorption lines; the bandpasses have
been chosen wide enough to enable a calibration of index dilution due to
velocity dispersions in galaxy nuclei. 
The indices have been defined using old Galactic stars and globular clusters,
free of emission lines. However, it is now known that a significant fraction
(at least 50\%) of `normal' elliptical galaxies contains ionized gas
(Goudfrooij \etal \cite{paul+94a} (hereafter GHJN) and references
therein).    
In this {\it Letter}, we study the effect of nebular emission lines to
measured line strengths. Particular attention will be devoted to the
effect of the \NI$\,\lambda\lambda 5197.9, 5200.4$ emission-line
doublet (often, and hereafter, refered to as \NI\,5199), since it is
situated right in the continuum bandpass redward of the \mgb\ lines,
and is thus a potential source of error.  

\section{Ionized gas in elliptical galaxies}
\label{gas}

Elliptical galaxies were originally thought to be essentially devoid of
gas. However, recent advances in instrumental sensitivity have caused a
renaissance of interest in the interstellar medium (ISM) of elliptical
galaxies. All known components of the ISM are now accessible using present-day
detector technology. 
In particular, GHJN found that more than 50\% of
a complete, optical magnitude-limited sample of giant elliptical galaxies from
the RSA catalog (Sandage \& Tammann \cite{rsa81}) exhibits \Ha+\NII\ emission. 
The emission-line regions are generally spatially extended, and associated
with dust lanes or patches. 

The emission-line spectra of giant elliptical galaxies are virtually always
similar to those of LINERs (Low-Ionization Nuclear Emission Regions, Heckman
\cite{heck80}) (GHJN; Goudfrooij \etal \cite{paul+94b}; Phillips
\etal \cite{pjdsb86}). 
Lines from low-excitation ions such as \NI\ are relatively strong in
LINERs. Median observed values and ranges of relative emission-line intensity
ratios of LINERs in the optical region are listed in Table~\ref{tab:relint};
the empirical LINER data are taken from Veilleux \& Osterbrock
(\cite{veiost87}), Kirhakos \& Phillips (\cite{kirphi89}), Osterbrock, Tran \&
Veilleux (\cite{ost+92}), and Ho, Filippenko \& Sargent (\cite{ho+93}). 
Mean equivalent widths (hereafter EWs) of the \NII$\,\lambda6583$ line in
early-type galaxies range up to $\sim\,$10~\AA\ (cf.\
Phillips \etal \cite{pjdsb86} and GHJN; the mean value of the
emission-line galaxies in the combination of their samples is 2.6 \AA).   
For a spectral energy distribution typical of a giant elliptical galaxy (e.g.,
Bica \& Alloin \cite{ba87}), this means that \NI\,5199 can have an EW up to
about 1.8 \AA\ in the centres of early-type galaxies. Since a typical value
for the EW of the \mgb\ line in nuclei of early-type galaxies is 5 \AA\ (e.g.,
Carollo \& Danziger \cite{cardan94}), the effect of the \NI\,5199 emission
line can be substantial.   

\section{Contamination of atomic and molecular indices by emission lines}
\label{simu}

Line strengths are derived from the flux in a central bandpass\footnote{We
use the notations of Gonz\'alez (\cite{gonz93}) in this Section.}
$[\lambda_{c1}, \lambda_{c2}]$ versus a local continuum level
interpolated from a set of bracketing bandpasses at either side
of the feature ($[\lambda_{b1}, \lambda_{b2}]$ 
and $[\lambda_{r1}, \lambda_{r2}]$). {\em Atomic} and {\em molecular}
indices can then be written, respectively: 
\[
\begin{array}{rcll}
I_a & = & \int_{\lambda_{c_1}}^{\lambda_{c_2}} \left( 1 - 
\frac{S\left( \lambda \right)}{C\left( \lambda \right)} \right)
\mbox{d}\lambda & \mbox{~~~~[\AA]} \\ [1.5ex] 
I_m & = & -2.5 \times \log_{10}{\frac{\int_{\lambda_{c_1}}^{\lambda_{c_2}} 
\frac{S\left( \lambda \right)}{C\left( \lambda \right)} \,
\mbox{{\scriptsize d}}\lambda}
{\lambda_{c_2} - \lambda_{c_1}}} & \mbox{~~~~[mag]}
\end{array}
\]
where $S(\lambda)$ is the original spectrum and $C(\lambda)$ the linearly 
interpolated continuum defined by:
\begin{eqnarray}
C\left( \lambda \right) & \equiv & S_b \cdot \frac{\lambda_r - \lambda}
{\lambda_r - \lambda_b} + S_r \cdot \frac{\lambda - \lambda_b}
{\lambda_r - \lambda_b} \;\;\;\; \mbox{where} \\
S_b & = & S\,[\lambda_{b1}, \lambda_{b2}], 
\;\; \lambda_b = \frac{\left(\lambda_{b_1}
+ \lambda_{b_2} \right)}{2} \\
S_r & = & S\,[\lambda_{r1}, \lambda_{r2}], 
\;\; \lambda_r = \frac{\left(\lambda_{r_1}
+ \lambda_{r_2} \right)}{2} 
\end{eqnarray}
where we have defined $F\,[\lambda_1, \lambda_2] \equiv
\int_{\lambda_1}^{\lambda_2} F(\lambda)\, \mbox{d}\lambda /
\left(\lambda_2 - \lambda_1\right)$ for a function $F=F(\lambda)$. We
will also use the central wavelength of the feature, $\lambda_c =
\left(\lambda_{c1} + \lambda_{c2}\right) / 2$.  

In the following, we will assume that the object spectrum $S^e(\lambda)$ 
is composed of a spectrum free of emission lines $S(\lambda)$ and an
emission-line spectrum $E(\lambda)$ so that $S^e(\lambda) = S(\lambda) +
E(\lambda)$.    

\subsection{Atomic indices} 

We can then derive an approximation for the difference in atomic
indices of $S^e(\lambda)$ and $S(\lambda)$:
\begin{eqnarray}
\Delta I_a & \simeq & - \frac{E[\lambda_{c1},\lambda_{c2}] 
\left(\lambda_{c2} - \lambda_{c1} \right)}{C(\lambda_c)} 
+ \frac{\lambda_{c2} - \lambda_{c1}}
{\lambda_r - \lambda_b} \times \nonumber  \\
&& \hspace*{-0.7cm} \times \left\{ \frac{E[\lambda_{b1},\lambda_{b2}] 
\left(\lambda_r - \lambda_c\right)}{S_b}
+ \frac{E[\lambda_{r1},\lambda_{r2}] \left(\lambda_c - \lambda_b\right)}{S_r}  
\right\}\mbox{.} \label{eq:Ia}
\end{eqnarray}
Here we assume that $S_r / S_b$ and $S(\lambda) / C(\lambda)$
are close to unity and that the emission-line intensity is small
compared to both $S_r$ and $S_b$, which is usually valid for a spectrum
dominated by stellar absorption lines. We checked that this formula
is a good approximation (to better than 10\%) by applying it to 
a number of stellar (spectral types G7 to M3) and galaxy spectra.

An emission line will usually only contribute to one of the
three bandpasses involved, leaving one non-zero term in Eq.~(\ref{eq:Ia}).
This is the case for e.g., the \NI\,5199 emission doublet which is located
inside the red bandpass of the \mgb\ index. 
Using the values of the limits for the \mgb\ bandpasses (Gonz\'alez
\cite{gonz93}), we find:  
\begin{equation}
\Delta \mbox{\mgb} \, \simeq \, 1.13 \times \left(
\frac{E[\lambda_{r1},\lambda_{r2}] \left(\lambda_{r2} - \lambda_{r1}
\right)}{S_r} \right)  
\label{eq:mgb} 
\end{equation}
The second term on the right-hand side of Eq.~(\ref{eq:mgb}) roughly
equals the EW of the emission line inside the wavelength range
$[\lambda_{r1},\lambda_{r2}]$. If the gas and stars have comparable
velocities and the width of the \NI\,5199 doublet is not too large ($\sigma <
300$~\kms), the major part of \NI\,5199 is emitted inside the red
continuum bandpass of the \mgb\ feature. For a typical EW of \NI\,5199 of
$0.5$ \AA, $\Delta \mbox{\mgb} \sim 
0.56$, but can be as large as $\sim 2$ for a nuclear spectrum of an
early-type galaxy (when EW of \NI\,5199~$\sim 1.8$~\AA, e.g., Emsellem
\etal \cite{eric+96}).  
To indicate the importance of the effect of \NI\,5199 to \mgb, we
note that the fraction of early-type galaxies with EW\,(\NI\,5199)
$\geq$ 0.5 \AA\ is $\sim 28$\% (combining the samples of Phillips \etal
\cite{pjdsb86} and GHJN).

The Hydrogen lines (e.g., \Ha\ and \Hb) present another problem,
since they are often present both in absorption (stellar) and in
emission (ionized gas). In this case, accurate correction for nebular emission
can only be achieved by carefully building a suitable absorption-line
template (\eg from early-type galaxies showing {\it no\/} evidence for
emission lines, cf.\ Section 4.2), and checking the template
subtraction procedure {\it a posteriori\/} by \eg comparing the kinematics
of the \Hb\ line with that of \NII\,6583. This will be detailed 
upon in a forthcoming paper. 

\subsection{Molecular indices}

We will deal here with the case when the emission line only contributes to
the central molecular feature.  If we again assume that the EW of the
emission line is small compared to the EW of the central feature, we
can write:  
\begin{equation}
\Delta I_m \, \simeq \, -1.086 \times \frac{EW(E)}
{\lambda_{c2} - \lambda_{c1} - EW(S)}
\label{eq:mol}
\end{equation}
where $EW(E)$ and $EW(S)$ are the equivalent widths of the emission line and
the molecular feature respectively, integrated inside the central bandpass. 

An example of emission-line contamination to molecular
indices is the effect of the \NI\,5199 doublet on the---very commonly
used---Mg$_2$ line strength. For a line width of 200~\kms\ 
and no difference of velocities between the gas and stars (i.e.,
$\Delta v = 0$), the EW of \NI\,5199 integrated inside the central
bandpass of the Mg$_2$ index ([5154.125~\AA, 5196.625~\AA]) represents
about 20\% of the total \NI\,5199 flux. Therefore, for a typical
\NI\,5199 equivalent width of $0.5$~\AA, $\Delta \mbox{Mg}_2$ would be
negligible.  
However, the contribution of the \NI\ doublet is higher when the gas has a
radial velocity smaller than that of the stars, which then displaces the
emission lines towards bluer wavelengths. This occurs quite often as shown
by Bertola \etal (\cite{bert+95}, hereafter B+95) who observed both the
stellar and gaseous components of a sample of S0 galaxies (e.g., 
counterrotating gas in NGC~4379, B+95). The velocity
difference can easily reach 100~\kms\ which increases the
contribution of \NI\,5199 into the Mg$_2$ central bandpass to $\sim 45$\%.
Then, for a strong \NI\,5199 line (EW = 1.8~\AA), we obtain
$\Delta \mbox{Mg}_2 \simeq 0.03$. These results obviously depend on
the spectral resolution which 
tends to dilute any feature into the continuum, and on the precise profile of
the emission-line in question (e.g., the ratio between \NI\,$\lambda5198$ and
\NI\,$\lambda5200$). 

\begin{table}[tb]
  \caption[]{Emission-line intensities of the central 2 arc\-sec of NGC~2974,
    corrected for reddening and relative to H$\beta$. Median values and       
    ranges for empirical LINER spectra are also given (see text).} 
  \begin{tabular*}{8.75cm}{@{\extracolsep{\fill}} l l c c l@{}}
  \hline \hline  
  \multicolumn{3}{c}{~~} \\ [-1.8ex]
  & & & \multi{2}{c}{LINER} \\
  Line & $\lambda_0$\,[\AA] & NGC~2974 & median & range \\ [0.5ex] \hline 
  \multicolumn{3}{c}{~~} \\ [-1.8ex]
    \OII\ & 3727 & 2.6 & 4.2 & 1.8$-$13 \\
  \NeIII\ & 3869 & 0.6 & 1.1 & 0.2$-$3.1 \\ 
     \Hb\ & 4861 & 1.0 & 1.0 & --- \\
   \OIII\ & 5007 & 1.6 & 1.4 & 0.3$-$3 \\
     \NI\ & 5199 & 0.6 & 0.3 & 0.1$-$0.7 \\
     \OI\ & 6300 & 0.8 & 0.9 & 0.2$-$4 \\
    \NII\ & 6548 & 2.1 & 1.3 & 0.6$-$4.6 \\
     \Ha\ & 6563 & 2.8 & 2.8 & --- \\
    \NII\ & 6583 & 6.3 & 3.9 & 1.7$-$14 \\
    \SII\ & 6724 & 3.4 & 2.8 & 1.1$-$7 \\
  \multi{2}{c}{~~} \\ [-1.8ex] \hline 
  \multi{2}{c}{~~} \\ [-1.5ex] 
  \end{tabular*}                                                              
  \parbox{8.5cm}{
  {\sl Notes:\/} \OII$\,\lambda3727$ stands for the $\lambda\lambda3726, 3729$
  doublet, \\ \NI$\,\lambda5199$ stands for the $\lambda\lambda5198, 5200$
  doublet, and \\ \SII$\,\lambda6724$ stands for the $\lambda\lambda6716,
  6731$ doublet.} 
  \label{tab:relint}
\end{table}
\section{Example: Mg\,$b$ and ionized gas in the inner region of NGC~2974} 
\label{example}

In this Section we illustrate the effect of \NI\,5199 to the \mgb\ index by
means of spectroscopy of NGC~2974, an elliptical galaxy featuring moderately
strong line emission in its inner region (GHJN). 

\subsection{Observations}
\label{obs}

A long-slit spectrum (resolution:\ 9 \AA\ {\sc fwhm}; wavelength interval:\ 
3650\,--\,7220 \AA) of NGC~2974 was
taken at the 1.52-m ESO telescope. We used the B\,\&\,C spectrograph,
and the total exposure time was 60 minutes. The slit was 
2 arcsec wide, and aligned with the photometric major axis of the galaxy.  
We also observed NGC~2974 with the TIGER 2-D spectrograph (Bacon \etal
\cite{baco+95}), mounted at the 3.6-m C.F.H.\ Telescope. We took 
several exposures representing a total of 120 minutes in the
wavelength interval 5100\,--\,5580 \AA\ to include the \mgb\ and Fe
absorption lines (hereafter refered to as the {\it absorption\/} data), and
exposures representing a total of 105 minutes were obtained in the wavelength
interval 6500\,--\,7000 \AA\ (hereafter the {\it emission\/}
data).  More detailed information on these observations of NGC~2974
and the data reduction will be described in a forthcoming paper
(Emsellem \& Goudfrooij \cite{ericpaul96}). 

\subsection{Spectral analysis}
\label{specanal}
\begin{figure} 
\psfig{figure=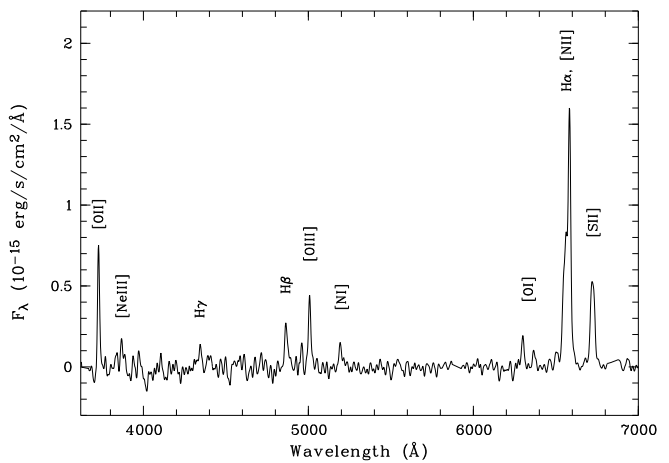,angle=-90,width=8.75cm}
\caption[]{The pure emission-line spectrum of the central 2 arcsec of
NGC~2974, after subtraction of a stellar absorption-line template (see text
for explanation). Prominent emission lines are labeled for reference}
\label{fig:spec2974} 
\end{figure}

To study the pure emission-line spectrum of the inner region of NGC~2974, we
used the method of Goudfrooij \etal (\cite{paul+94a}), which consists of
building a best-fit absorption-line template by averaging nuclear spectra of
elliptical galaxies which do not have any evidence for emission lines, after
correction for differences in redshift and velocity dispersion (see Emsellem \&
Goudfrooij \cite{ericpaul96} for a more detailed description).
Prominent telluric absorption 
lines were interpolated over. The resulting spectrum of the central 2 arcsec
is shown in Fig.~\ref{fig:spec2974}. Emission-line fluxes were measured 
using gaussian fits, and listed in Table~\ref{tab:relint}. The line-intensity
ratios are consistent with those of LINERs. 

\begin{figure}[tb]
\psfig{figure=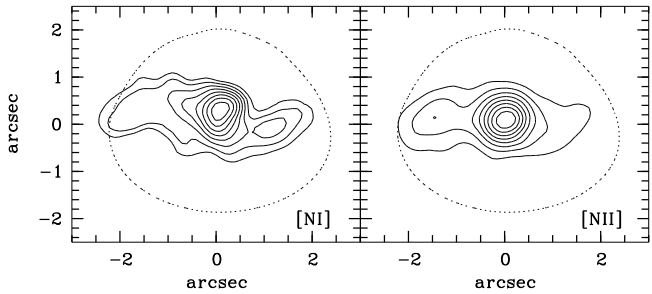,angle=0,width=8.75cm}
\caption[]{Two-dimensional contour maps of the emission-line flux 
in NGC~2974: \NI\,$\lambda$5199 (left panel) and \NII\,$\lambda$6583
(right panel). The contours correspond to levels from 20\% to 90\% of
the maxima with a step of 10\%. An isophote of the stellar continuum
(dashed line) is shown for comparison}
\label{fig:map_NI} 
\end{figure}

\begin{figure} 
\psfig{figure=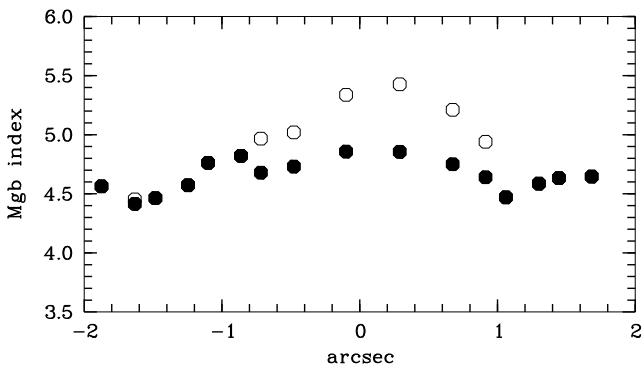,angle=0,width=8.75cm}
\caption[]{Minor-axis profile of the \mgb\ index in NGC~2974 from the 
TIGER spectra:\ line strength before (open symbols) and after (filled
symbols) correction for the \NI\,$\lambda$5199 emission}
\label{fig:mgb}
\end{figure}

Taking advantage of the two-dimensional coverage of the TIGER spectrograph,
we mapped the flux distribution of each individual emission line.
We indeed detected the \NI\,5199 emission lines in the TIGER {\it absorption}
data:\ its map is presented in Fig.~\ref{fig:map_NI} together with the
one of the \NII\,6583 line. 
Although the uncertainty on the {\it absolute\/} \NI\,5199 flux
is non-negligible, the similarity of these maps is striking, especially 
since the stellar continuum is rather featureless (except for some
dust absorption along the major-axis). The gas kinematics derived from
the \NI\,5199 and \NII\,6583 lines are also essentially identical (Emsellem \&
Goudfrooij \cite{ericpaul96}). We therefore argue that the 
detected \NI\,5199 emission is real and not due to any spurious effect
(e.g., the spectral fitting process).
We then derive the \mgb\ line strength from the original TIGER
absorption spectra, and from the corresponding spectra from which the 
fitted \NI\ emission has been subtracted 
%
%
(see Emsellem \etal \cite{eric+96} for details). 
The measured difference $\Delta \mbox{\mgb}$ is large in the centre,
up to $\sim 0.9$~\AA. In Fig.~\ref{fig:mgb}, we present the
corresponding minor-axis profiles which show that the central \mgb\
enhancement nearly disappears after the correction for \NI\,5199 emission. 

\section{Conclusions}
\label{concl}

In this {\it Letter}, we have discussed the effect of nebular emission lines
on the derivation of absorption-line strengths in the central region
of early-type galaxies. We derived analytical expressions 
for changes induced by the presence of nebular emission lines, both
for atomic and molecular absorption-line indices. 
In particular, we emphasize the fact
that the measurement of the well-known \mgb\ index can be strongly perturbed
by the presence of the \NI\,5199 doublet, with a difference in equivalent
width $\Delta \mbox{\mgb}$ up to 2~\AA. 
We have illustrated this effect using spectrographic observations
of NGC~2974. We argue that this effect should be considered when studying 
line-strength gradients in the inner regions of early-type galaxies.

\acknowledgements{
We thank the anonymous referee for a thorough reading of the
manuscript. We have made use of the NASA/IPAC Extragalactic Database 
(NED) which is operated by the Jet Propulsion Laboratory, Caltech, 
under contract with the National Aeronautics and Space Administration.
}

\end{document}